\documentclass[journal=jcisd8,manuscript=article,layout=twocolumn]{achemso}
\setkeys{acs}{articletitle=true, maxauthors=5}

\usepackage{svg}
\usepackage{lineno}
\usepackage{multicol}
\usepackage[font=footnotesize,labelfont=bf]{caption}
\usepackage{subcaption}
\usepackage[hidelinks]{hyperref} 
\usepackage{pifont} 
\usepackage[utf8]{inputenc}
\usepackage{amsmath, amsfonts}
\usepackage{multirow, array, threeparttable, adjustbox, hhline}
\usepackage{bm}                   
\usepackage{xcolor}   

\definecolor{p1}{HTML}{103240} 
\definecolor{p2}{HTML}{1D6373} 
\definecolor{p3}{HTML}{4995A6} 
\definecolor{p4}{HTML}{732929} 
\definecolor{p5}{HTML}{A64141} 

\newcommand{\eg}{\textit{e.g. }}

\newcommand{\etal}{\textit{et al. }}

\author{Leonid Komissarov}
\affiliation[CCM]{Center for Molecular Modeling (CMM), Ghent University, Technologiepark-Zwijnaarde 46, B-9052, Ghent, Belgium}
\alsoaffiliation[SCM]{Software for Chemistry \& Materials (SCM) B.V., De Boelelaan 1083, 1081 HV Amsterdam, The Netherlands}

\author{Robert Rüger}
\affiliation[SCM]{Software for Chemistry \& Materials (SCM) B.V., De Boelelaan 1083, 1081 HV Amsterdam, The Netherlands}

\author{Matti Hellström}
\affiliation[SCM]{Software for Chemistry \& Materials (SCM) B.V., De Boelelaan 1083, 1081 HV Amsterdam, The Netherlands}

\author{Toon Verstraelen}
\affiliation[CCM]{Center for Molecular Modeling (CMM), Ghent University, Technologiepark-Zwijnaarde 46, B-9052, Ghent, Belgium}
\email{toon.verstraelen@ugent.be}
\abbreviations{}

\title{\textbf{ParAMS: Parameter Optimization for Atomistic and Molecular Simulations}}

\begin{document}

\begin{abstract}
This work introduces ParAMS -- a versatile
Python package that aims to make parameterization workflows
in computational chemistry and physics more accessible, transparent and reproducible.
We demonstrate how ParAMS facilitates the parameter optimization for potential energy surface (PES) models, which can otherwise be a tedious specialist task.
Because of the package's modular structure, various functionality can be easily combined to implement a diversity of parameter optimization protocols.
For example, the choice of PES model and the parameter optimization algorithm can be selected independently.
An illustration of ParAMS' strengths is provided in two case studies:
i) a density functional-based tight binding (DFTB) repulsive potential for the inorganic ionic crystal ZnO, and 
ii) a ReaxFF force field for the simulation of organic disulfides.
\end{abstract}

\newpage
\section{Introduction} \label{sec:intro}
Throughout the years,
the use of predictive computational models has become standard practice
in many professional fields such as logistics, economics or R\&D
\cite{generic0, generic1, generic2, generic3, generic4}, with computational
chemistry being no exception.
Predictive models in this field often approximate
the potential energy surface (PES) and derived properties for a given chemical system,
and can be broadly categorized based on their level of theory:
Quantum mechanical (QM) approaches
such as wave function or density functional theory
explicitly model the electronic structure,
generally resulting in a high accuracy and broad applicability.
However, explicit electronic treatments come with a high computational price tag,
making calculations unfeasible for larger systems.
This limitation can be overcome by introducing more empiricism to the model
on the electronic (\eg tight binding), atomic (\eg molecular mechanics) or 
molecular (\eg coarse-graining)
level.
Such empirical models attempt to strike the balance between speed and prediction accuracy by approximating the PES description through the introduction of parameters.
Prominent examples include approximate functionals in
density functional theory (DFT) \cite{becke3, lyp, b3lyp, wB97XD},
density functional tight binding (DFTB) \cite{dftb1, dftb2, dftb3, xtb1, xtb2},
machine learning (ML) potentials \cite{behler, schnet, ani1x, pinn},
and force fields (FF). \cite{amber, charmm, reax1, reax2}

While some of the best empirical models can closely approximate higher-level QM theories
at only a fraction of the computational time \cite{ani1ccx, gfn2xtb},
their quality strongly varies with different sets of parameters
\cite{anna,ogolem,forcebalance,potfit,garfield}.
Additionally, many empirical models can only deliver accurate results for a comparably limited chemical space,
either due to their functional form or parameters being specific to certain (combinations of) elements.
These limitations can lead to the existence of multiple parameter sets
based on the desired application, \eg ReaxFF has distinct families of combustion or condensed phase parameterizations. \cite{reax_npj}
Such lack of general parameters gives rise to the research field of parameter fitting
\cite{fitting1, fitting2, fitting3, fitting4, anna},
where the task is to find an optimal parameter set, given training data
constructed from chemical systems and their properties of interest.
Although the fitting process can be an appealing solution to the above shortcomings,
its practical implementation remains hardly accessible to the broader audience
and instead is almost exclusively carried out by specialized research groups.
In our experience, the majority of researchers,
although being interested in individual parameter fitting,
are discouraged by the high barrier that comes with it.
The main reason for this being a lack of generalization and transparency:
Training data often comes in a variety of formats, optimizers expect a different input all together and the format in which parameters are stored is specific to each method. \cite{reax_manual, forcebalance_manual, ogolem_manual, posmat}
The combination of these oftentimes results in works that can be hardly comprehended and reproduced by third parties.
In an effort to address the above issues, we introduce the ParAMS scripting package for Python. 
The following section briefly summarizes the architecture of ParAMS and we refer to the documentation for further technical details. \cite{doc}
The Results section demonstrates how the package can be used to i) generate density functional-based tight binding (DFTB) two-body repulsive potentials for an ionic material, and ii) reparameterize a ReaxFF reactive force field \cite{reax1, reax2} for organic disulfides.
Step-by-step Jupyter notebooks for the two case studies are provided as supporting information, as well as on GitHub \cite{si}.
Additional application examples are available in the package's documentation. \cite{doc}
The final section concludes with a summary and an outlook on future work.

\section{Implementation}
ParAMS follows a modular package structure
with well-defined application programming interfaces (APIs),
which allows components to be treated independently.
This is essential for future development, as individual sub-modules can
be easily worked on and extended.
We describe the main components and their functionality below.
For a mathematical description of the functionality as well as 
additional explanation of the syntax, please refer to section S1 of the supporting information.

\textbf{Job Collection and Data Set}
classes are responsible for the input / output (IO) of relevant data.
In the context of ParAMS, these will be collections of chemical systems, properties and settings alongside optional metadata.
A Job Collection clearly defines job entries by combining systems and settings.
The Data Set defines which properties of a job are relevant to the optimization and stores
the reference values of all entries in a vector $\bm{y}$.
Reference values can be added from any source: Experimental / external results, or a high-level calculation.
To ensure reproducibility and ease-of-use, ParAMS makes use of the
YAML data-serialization format \cite{yaml} as the default for all IO operations.

\textbf{Extractors} tell ParAMS how to 
extract a property of interest $P$ from a calculated job.
Technically, they are small standalone Python modules that read the native
(\eg stream, text or binary file) output of the Amsterdam Modeling Suite\cite{ams, scm} into Python variables.
Examples for implemented extractors are:
Interatomic distances, valence angles, dihedral angles, atomic charges, reaction energies,
linear transit energy profiles, lattice vectors, bulk moduli, atomic forces, stress tensors, Hessian matrices and vibrational frequencies.
New extractors are easily written by the user,
effectively allowing any  property that can be calculated with AMS
to be fitted within the scope of ParAMS.
Extractors also support more elaborate cases that require additional processing before a comparison. This is for example needed when computing the minimal root-mean-square deviation of atomic positions.

\textbf{Loss Functions} implement various metrics that describe the distance between two vectors.
In the context of parameter fitting, a vector consisting of all reference values 
$\bm{y}$ has to be compared to the predictions vector $\bm{\hat{y}}$,
as generated given a specific set of parameters,
in order to measure the quality of the fit.
Implemented metrics are least absolute error (LAE), mean absolute error (MAE),
root-mean-square error (RMSE) and residual sum of squares (RSS).
Additionally, user-defined metrics are supported.

\textbf{Optimizers} provide a unified interface to a variety of optimization algorithms.
Currently, the following are supported:
Covariance Matrix Adaptation Evolution Strategy (CMA-ES) \cite{cma1,cma2,cma_py},
Adaptive Rate Monte Carlo (ARMC) \cite{armc} and
optimizers available through the Nevergrad \cite{nevergrad} and SciPy \cite{scipy} packages. 
To guarantee parameterization support for a wider range of models,
the current version of ParAMS is designed to work with gradient-free optimization algorithms only.

\textbf{Parameter interfaces} translate the parameter vector $\bm{x}$ into the native
format of the empirical model (\eg a file on disk).
Any existing parameter interface can be parameterized.
At the time of writing, ParAMS supports interfaces to
ReaxFF \cite{reax_ADF}, SCC-DFTB repulsive potentials, GFN1-xTB \cite{xtb1}, and  Lennard-Jones potentials.

\textbf{Callbacks} allow the interaction with a parameterization at runtime.
Such interactions can be progress loggers, timeouts, early stopping criteria or plotting functions \cite{matplotlib}.

The \textbf{main script} is a command line interface to ParAMS for users who do not wish
to spend much time writing their own parameterization scripts.
It allows the setup and execution of the most common tasks through a configuration file.

\begin{figure}
 \centering
 \includegraphics[width=.4\textwidth]{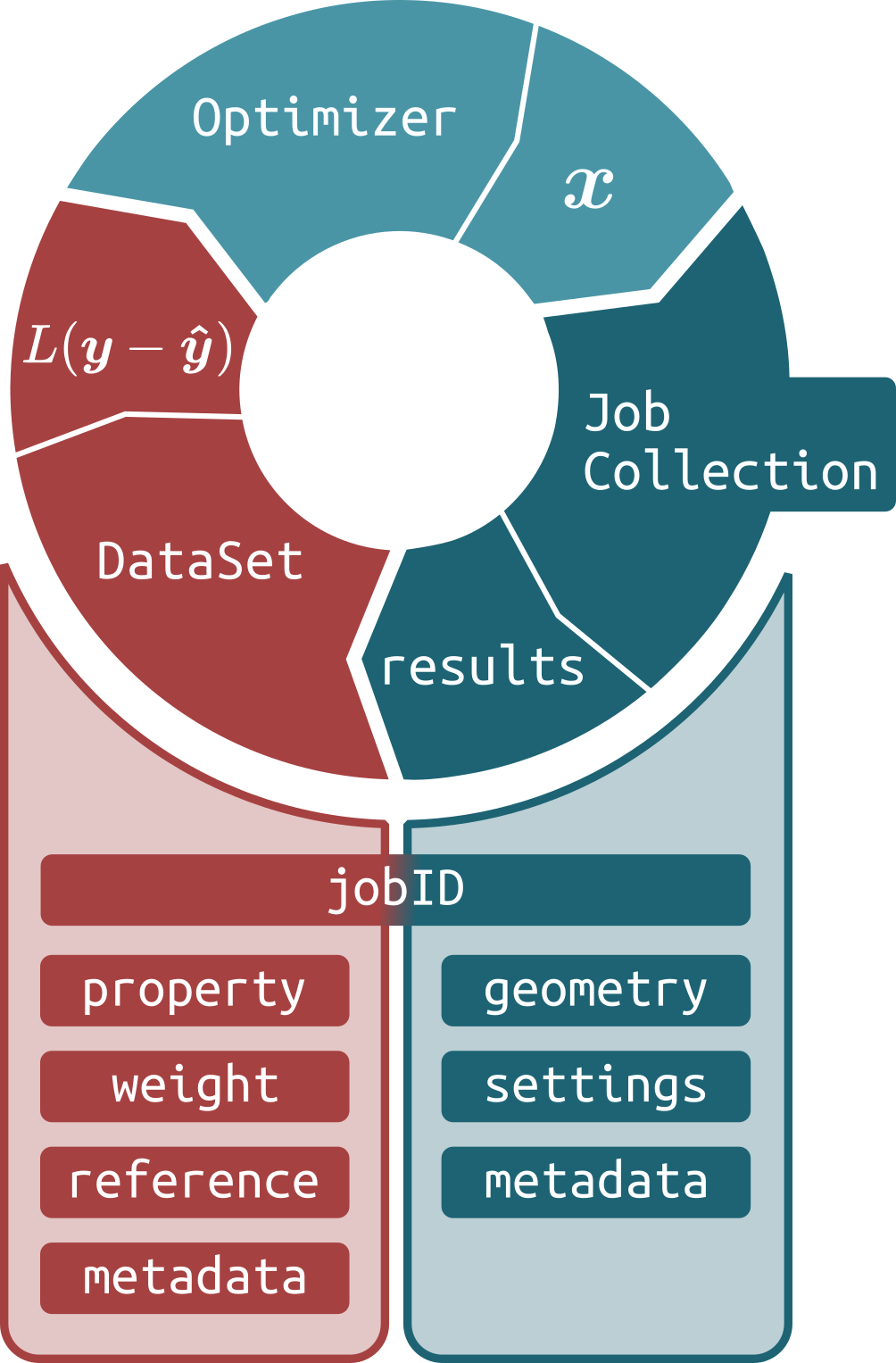}
 \caption{
 A schematic representation of the ParAMS parameterization loop.
 Highlighting the interplay between the three main components:
 Optimizer, Job Collection and Data Set.
 Every parameter set suggested by the optimizer produces different job results,
 which are evaluated by the Data Set based on a unique \textit{jobID}.
 The results are then compared with the reference values,
 weighted and combined into the loss function value.
 A more detailed description is available in the package documentation.
 }
 \label{fig:loop}
\end{figure}

Figure \ref{fig:loop} shows the general parameterization loop and highlights the main input-output relationships.
The implementation allows users already experienced in other data science packages to prepare inputs and process results in a familiar way.
Techniques like training- and validation set splitting,
cross-validation, early stopping, batching or outlier detection are supported out of the box.
Parameter constraints can be further used to limit the search space of a problem.
In addition to regular box constraints,
users can express inequality constraints involving multiple parameters (\eg $x_1 \leq x_2$), as will be demonstrated in the ReaxFF example application.
ParAMS implements two levels of parallelism.
Multiple parameter vectors and multiple jobs per vector can be evaluated at the same time,
resulting in workloads that can be distributed effectively based on the training set size and optimization algorithm.
Moreover, since the signatures of all classes in a submodule are the same,
different models and optimizers can be effortlessly compared and deployed (\eg comparison of different levels of theory in Tab. \ref{tab:repulsive}).
Daily regression tests are performed to guarantee an error-free functionality of the package.

\section{Results}
\subsection{DFTB two-body repulsive potential parameterization}\label{sec:dftb}
Here, we illustrate how the ParAMS package can be used to train a two-body SCC-DFTB repulsive potential \cite{dftb1}. 
For simplicity, we use ZnO, for which several previous parameterizations already exist in the literature \cite{znopt}. We will approximately follow the approach from Ref. \citenum{znopt}, in which the authors reused the electronic parameters from the znorg-0-1 DFTB parameter set \cite{znorg} and reparametrized the two-body Zn-O repulsive potential to reference data calculated for the wurtzite and rocksalt polymorphs of ZnO. With the znorg-0-1 parameters, the rocksalt polymorph is predicted to be more stable than the wurtzite polymorph, but in experiments and DFT calculations, the opposite is true, which motivates the reparameterization.

The entire code needed for the parameterization is provided as supporting information \cite{si}. It fits the Zn-O pairwise repulsive potential $V^\text{rep}$ as a tapered double exponential function of the form
\begin{equation}
\begin{split}
    &V^\mathrm{rep}(r) = \\ &[A_0\exp(-A_1 r) + A_2\exp(-A_3 r)]f^\mathrm{cut}(r) 
\end{split}
\end{equation}
where $A_0$, $A_1$, $A_2$, $A_3$ are the parameters and $f^\text{cut}(r)$ is a tapering function of the form
\begin{equation}
    f^\text{cut}(r) = \frac{1}{2}\left(\cos(\frac{\pi r}{r_\text{cut}})+1\right)
\end{equation}
with the cutoff distance $r_\text{cut} = 5.67$ bohr.

With ParAMS, it is possible to either directly define the reference values for the training set (for example, from literature values) or to automatically calculate them if no reference values have been given. Here, we illustrate the second approach, and perform the reference calculations using the periodic DFT code BAND\cite{band} in the Amsterdam Modeling Suite\cite{ams} (AMS). 

The training set comprises the $a$ and $c$ lattice parameters of wurtzite ZnO, the bulk modulus $B_0$ of wurtzite ZnO, as well as the relative energies of the wurtzite and rocksalt polymorphs of ZnO,
$\Delta E = E_\text{wurtzite}-E_\text{rocksalt}$ (per ZnO formula unit). We do not need to specify the reference values themselves, as they will automatically be calculated by ParAMS.

The job collection contains two jobs: lattice optimizations of the wurtzite and rocksalt polymorphs. From these two jobs, the $a$, $c$, $B_0$, and $\Delta E$ quantities can be extracted. For wurtzite, $B_0$ can be extracted by requesting that the elastic tensor be calculated at the end of the lattice optimization.

The reference DFT calculations were run with the PBE exchange-correlation functional, a triple-$\zeta$ (TZP) basis set, and ``Good'' numerical quality (dense k-space and integration grids). For the parametrized DFTB engine, a ``Good'' (dense) k-space grid was also used, since the results of lattice optimizations can be quite sensitive to the k-space grid.

\begin{figure}[htb]
    \centering
    \includegraphics[width=0.5\textwidth]{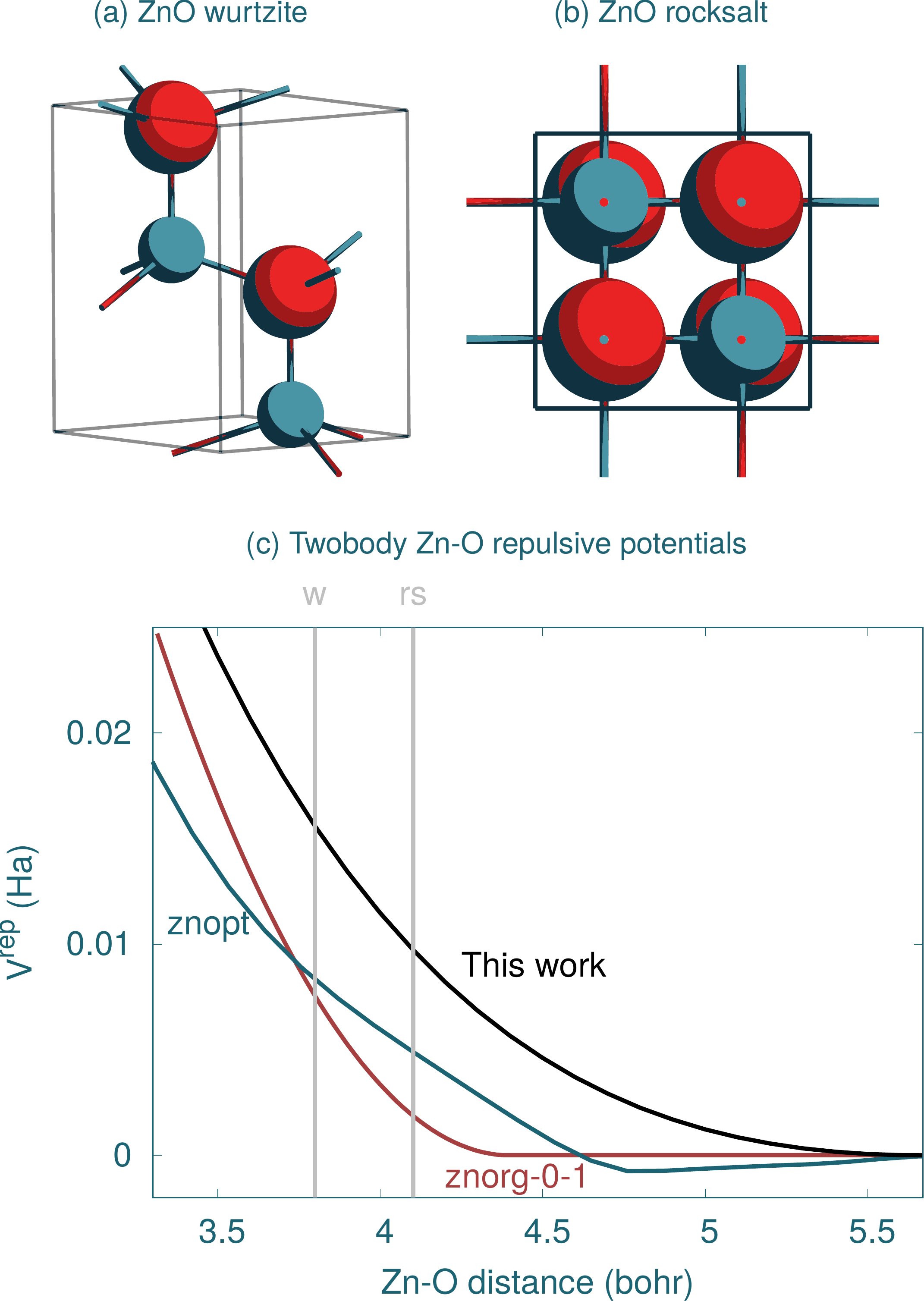}
    \caption{
    ZnO parameterization data.
    (a-b) pictures of the unit cells of ZnO wurtzite and rocksalt polymorphs. Zn is shown in blue and O in red. (c) DFTB Zn-O pairwise repulsive potentials for the znorg-0-1 \cite{znorg} (red), znopt \cite{znopt} (blue), and refitted repulsive potential from this work (black). The gray lines mark typical Zn-O distances in the wurtzite (w) and rocksalt (rs) polymorphs, respectively.} 
    \label{fig:repulsive}
\end{figure}

The optimization was done with the Nelder-Mead algorithm \cite{neldermead} from scipy, with a sum-of-squared-errors loss function. The smallest loss function value was obtained for $A_0 = 0.45$, $A_1 = 1.01$, $A_2 = 0.25$, $A_3 = 0.40$. The resulting repulsive potential is shown in Figure \ref{fig:repulsive}. Typical Zn-O distances in wurtzite (``w") and rocksalt (``rs") are indicated with gray lines at 3.8 bohr and 4.1 bohr, respectively. With znorg-0-1 (red line), the repulsive potential decays very rapidly between the typical wurtzite and rocksalt distances. This decrease is not as pronounced with either the potential in this work (black line) or znopt (blue line). This affects the relative stability of the wurtzite and rocksalt ZnO polymorphs. 

Table \ref{tab:repulsive} compares the resulting ZnO properties from the parameterization in this work to the DFT reference data, as well as previous DFTB ZnO parameterizations (note: znorg-0-1 was not parametrized to the DFT data in Table~\ref{tab:repulsive}, and znopt was parametrized to also reproduce some adsorption energies). The repulsive potential in this work closely reproduces the wurtzite lattice parameters $a$ and $c$ and the relative energy $\Delta E$, and provides a good estimate of the bulk modulus, compared to the DFT reference to which it was trained. 

With ParAMS it is additionally possible to evaluate the loss function, and individual training set entries, using any of the engines in the Amsterdam Modeling Suite. For comparison, Table~\ref{tab:repulsive} also gives the corresponding quantities for the UFF (Universal Force Field) engine, which performs significantly worse than the DFTB parameterizations for these quantities.

\begin{table*}
    \centering
    \begin{tabular}{l|llll}
         & $a$ (\AA) & $c$ (\AA) & $B_0$ (GPa) & $\Delta E$ (eV) \\
         \hline
         DFT (this work) & 3.30 & 5.32 & 126 & $-$0.24 \\
         DFTB (this work) & 3.28 & 5.34 & 146 & $-$0.24 \\
         DFT \cite{znopt} & 3.29 & 5.31 & 129 & $-$0.30 \\
         DFTB, znopt \cite{znopt} & 3.21 & 5.25 & 161 & $-$0.32 \\
         DFTB, znorg-0-1 \cite{znorg} & 3.29 & 5.38 & 161 & +0.14 \\
         UFF & 2.90 & 4.73 & 199 & $-$16.2 \\
    \end{tabular}
    \caption{Calculated ZnO wurtzite lattice constants $a$ and $c$, bulk modulus $B_0$, and energy relative to the rocksalt polymorph $\Delta E = E_\text{wurtzite}-E_\text{rocsksalt}$ (per ZnO formula unit)}
    \label{tab:repulsive}
\end{table*}

\subsection{ReaxFF parameterization} \label{sec:application}
ReaxFF is another contemporary example of an empirical model. \cite{reax1,reax2}.
This formalism has been applied to a wide range of chemical problems,
and consequently has seen a lot of new parameter development
\cite{reax_RDX, reax_Silica, muellerhartke}
(for a general overview of ReaxFF and its development, see Senftle \textit{et al.}\cite{reax_npj}).
In this section, we demonstrate the parameterization of ReaxFF with a training set
previously published by Müller and Hartke (MH) \cite{muellerhartke}. The optimized
parameter vector $\bm{x}^*$, as found by MH, is called Mue2016.
An overview of the training set can be found in Table S1 of the supporting information.
It features a total of 231 geometries needed for the computation of 4875 chemical properties. Additionally, MH included a validation set to check for overfitting.
Examples of three structures included in the data are depicted in Fig. S1 of the supporting information, showing cyclopentathione, diphenyl disulfide and 
dimethyl disulfide.
Prior to the parameter optimization, we evaluate
the training and validation sets with the Mue2016 ReaxFF parameters and
report sum-of-squared-errors (SSE) losses of 14441 and 14451 respectively.
Note that in the original publication,  MH report a training set
loss of 12400\cite{muellerhartke}, while in a more recent work, Shchygol \etal calculate a loss of 16300\cite{anna}.
Such differences are expected, because numerical instabilities inherent to ReaxFF and software improvements (mostly related to geometry optimization) may result in different optimized geometries. \cite{anna, tapered_bond_orders}

In our setup, we use the covariance matrix adaptation evolution strategy (CMA-ES)
\cite{cma1, cma2, cma_py} as the optimization algorithm with Mue2016 as the initial point.
CMA-ES is gradient-free, and relies on a population to sample
new parameter vectors from an adapted, n-dimensional Gaussian.
It does not require additional hyperparameters other than a population size and an initial
width of the Gaussian distribution, $\sigma$.
Here, we use a population size of 36 and an initial $\sigma$
of 0.3.
Furthermore, we limit the optimization to 24 hours and set up an 
early stopping mechanism based on the validation set.
The optimization is set up to stop early only if there has been no
improvement in the validation set error for the last 6000 evaluations.

Rather than optimizing the same 87 parameters as MH,
we perform a one-dimensional scan on all
parameters and select the 35 most sensitive with respect to the training set:
Although Mue2016 is a set of 701 parameters in total,
only a subset of these significantly affects the overall cost function value.
This is for example the case when a model includes parameters for each chemical element (\eg C, H, O),
but the total training set of systems $R$ can be constructed from fewer elements (\eg C, H).
In such cases, the dimensionality of the problem can be reduced by scanning for a
relevant parameter subset which yields the biggest change
in the cost function value.
The simplest setting, which we used in this case study, only modifies one parameter at a time to determine its influence on the objective function.
It is also possible to scan all parameter combinations, to discover coupling between parameters, albeit at a highly increased computational cost.
Out of the 35 parameters selected this way, 16 have also been optimized by MH.
We list all optimized parameters in the provided Python notebooks\cite{si}.

Parameter bounds are set to be relative to the initial values such that 
$\bm{x}_\pm = \bm{x}_0 \pm 0.2|\bm{x}_0|$.
In addition to box constraints, ParAMS enables a definition of inequality constraints.
As the ReaxFF formalism works with bond orders, we limit the parameters responsible for the covalent radii of
$\sigma$, $\pi$ and $\pi\pi$ bonds to
$r_0^\sigma \geq r_0^\pi \geq r_0^{\pi\pi}$
for every atom and atom pair defined in the force field.
This approach effectively limits the search space and is available in combination with all optimizers.

\begin{figure}
  \includegraphics[width=.5\textwidth]{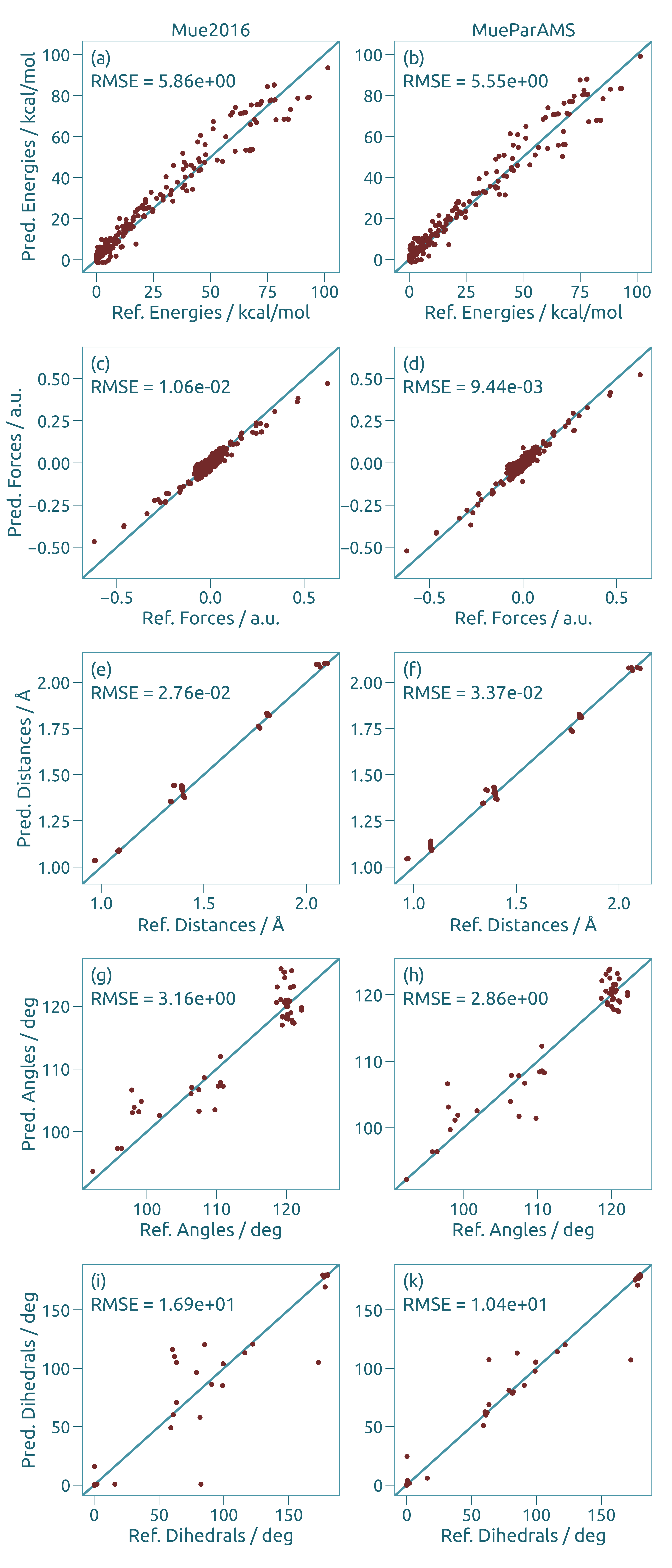}
  \centering
  \caption{
    Training data correlation plots. Showing
    (a,b) energy differences, (c,d) atomic forces, (e,f) atomic distances, (g,h) interatomic angles and (i,k) dihedral angles
    as calculated with the Mue2016 (left) and MueParAMS (right) parameters. Reaction energies and internal coordinates are compared after geometry optimization.
    X and Y axes depict reference and predicted values respectively.
    }
  \label{fig:corrplot}
\end{figure}

\begin{figure}
  \centering
  \includegraphics[width=.45\textwidth]{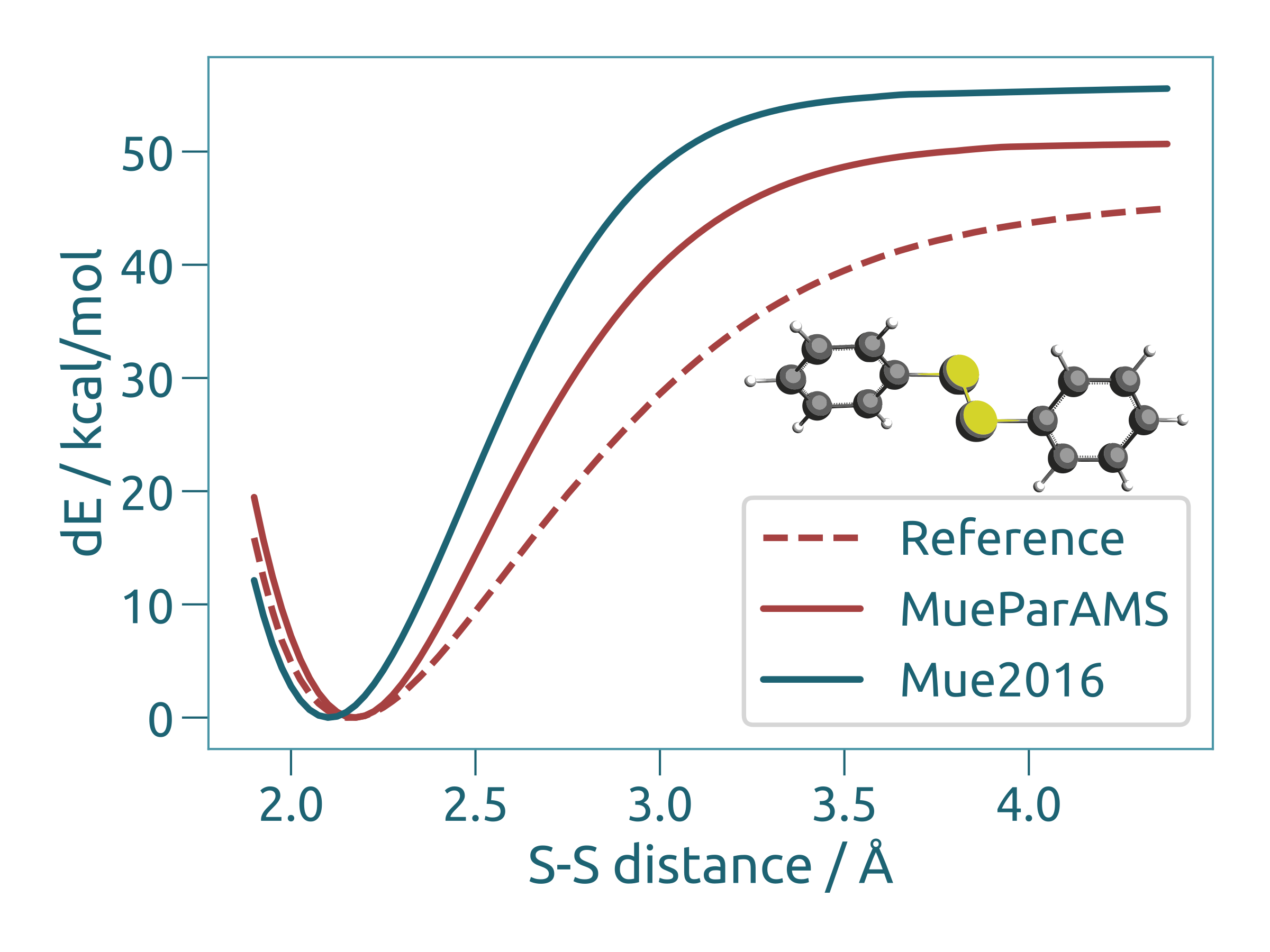}
  \caption{
    S-S dissociation curve of diphenyl disulfide. Computed with the
    Mue2016 and MueParAMS ReaxFF parameterizations. For details about the reference data, see Ref. \citenum{muellerhartke}.
    }
  \label{fig:pes}
\end{figure}

A summary of all settings is provided in Table S2 in the supporting information.
To compensate for the randomness of CMA-ES, we repeat the optimization
set-up nine times.
For the best solution, we report improved training and validation set losses of 11877 and 5377 respectively.
We make this work's optimized parameter set available through the supporting information under the title MueParAMS.
Correlation plots between reference and predicted values for the
new parameters are presented in Figure \ref{fig:corrplot}, showing very good agreement to the reference data.
Moreover,
Figure \ref{fig:pes} compares the S-S dissociation curve of
diphenyl disulfide, as computed with Mue2016 and the new MueParAMS parameters, showing an improved agreement to the reference data for this case.

\section{Summary and Outlook}
With ParAMS, we have presented a modern Python package, supporting versatile parameterization workflows with minimal effort.
Its integration with the Amsterdam Modeling Suite adds a high amount of
flexibility through the number of properties that can be fitted
alongside the support for multiple codes when it comes to the model, optimization algorithm and reference data selection.
Features such as highly customizable optimizations,
support for multiple validation sets, or the
intuitive processing of data
aim to make ParAMS accessible to both, advanced and less experienced users.
At the same time, developers can easily extend existing functionality.
We showed how an SCC-DFTB repulsive potential could easily be parameterized for the inorganic crystal ZnO. The reference data was calculated automatically using a DFT engine within AMS.
This example application also demonstrates how ParAMS can be used
to compare the accuracy of different chemical simulation packages given the same
training set.
We also demonstrated how the package can be used to easily process, set up and start a fitting procedure for ReaxFF.
Using previously published data by Müller and Hartke,
we were able to find parameters that produce a considerably lower error for the validation set
while maintaining a similar accuracy in the training data.
In future, we hope to extend the number of empirical models that can be fitted with ParAMS
and further improve the ease of use through the introduction of additional shortcut functions for training set building.
We also expect additions in other functionality such as optimization algorithms or extractors
based on user feedback and wishes as the project matures.
The package is included in all AMS releases since 2020.

\textbf{Supporting Information Available}
PDF file with mathematical description of the optimization problem, summary of the reference data published by Müller and Hartke used in the ReaxFF example, settings used for the ReaxFF parameterization and visualization of some structures in the Müller and Hartke training set.

\textbf{Data and Software Availability}
All data needed to reproduce the examples is available
at \url{www.doi.org/10.5281/zenodo.4629706}.
The package's documentation is available at
\url{www.scm.com/doc.trunk/params}.
ParAMS is distributed with the Amsterdam Modeling Suite, for which a free trial can be requested at \url{www.scm.com}.

\begin{acknowledgement}
We thank Michał Handzlik for the initial design of the 
ParAMS package and Dr.\ Tomáš Trnka for the implementation of the AMSWorker interface,
resulting in a considerable computational speed-up.
This project has received funding from the European Union's Horizon
2020 research and innovation programme under grant agreement No 814143 (L.K. \& T.V.), No 798129 (M.H).
T.V. also acknowledges funding of the research board of Ghent University.
R.R. and M.H. have received funding from the Netherlands Enterprise Agency (RVO) and Stimulus under the MIT R\&D Collaboration programme, project number PROJ-02612.
The computational resources (Stevin Supercomputer Infrastructure) and services used in this work were provided by the VSC (Flemish Supercomputer Center), funded by Ghent University, FWO and the Flemish Government – department EWI.
\end{acknowledgement}

\textbf{Author Contributions}
L.K. and R.R. developed the ParAMS Python package.
L.K. and M.H. performed the case studies.
L.K., M.H. and T.V. wrote the paper.
T.V. oversaw the project.
All authors read and approved the final manuscript.

\textbf{Competing Interests}
Authors L.K., R.R. and M.H. were employed by the company Software for Chemistry and Materials (SCM). SCM develops and commercializes the Amsterdam Modeling Suite, of which ParAMS is a new module.

\footnotesize{
\bibliography{references}

\providecommand{\latin}[1]{#1}
\makeatletter
\providecommand{\doi}
  {\begingroup\let\do\@makeother\dospecials
  \catcode`\{=1 \catcode`\}=2 \doi@aux}
\providecommand{\doi@aux}[1]{\endgroup\texttt{#1}}
\makeatother
\providecommand*\mcitethebibliography{\thebibliography}
\csname @ifundefined\endcsname{endmcitethebibliography}
  {\let\endmcitethebibliography\endthebibliography}{}
\begin{mcitethebibliography}{60}
\providecommand*\natexlab[1]{#1}
\providecommand*\mciteSetBstSublistMode[1]{}
\providecommand*\mciteSetBstMaxWidthForm[2]{}
\providecommand*\mciteBstWouldAddEndPuncttrue
  {\def\EndOfBibitem{\unskip.}}
\providecommand*\mciteBstWouldAddEndPunctfalse
  {\let\EndOfBibitem\relax}
\providecommand*\mciteSetBstMidEndSepPunct[3]{}
\providecommand*\mciteSetBstSublistLabelBeginEnd[3]{}
\providecommand*\EndOfBibitem{}
\mciteSetBstSublistMode{f}
\mciteSetBstMaxWidthForm{subitem}{(\alph{mcitesubitemcount})}
\mciteSetBstSublistLabelBeginEnd
  {\mcitemaxwidthsubitemform\space}
  {\relax}
  {\relax}

\bibitem[Bruzda(2018)]{generic0}
Bruzda,~J. Multistep quantile forecasts for supply chain and logistics
  operations: bootstrapping, the {GARCH} model and quantile regression based
  approaches. \emph{Cent. Eur. J. Oper. Res.} \textbf{2018}, \emph{28},
  309--336\relax
\mciteBstWouldAddEndPuncttrue
\mciteSetBstMidEndSepPunct{\mcitedefaultmidpunct}
{\mcitedefaultendpunct}{\mcitedefaultseppunct}\relax
\EndOfBibitem
\bibitem[Wolfers and Zitzewitz(2004)Wolfers, and Zitzewitz]{generic1}
Wolfers,~J.; Zitzewitz,~E. Prediction Markets. \emph{J. Econ. Perspect.}
  \textbf{2004}, \emph{18}, 107--126\relax
\mciteBstWouldAddEndPuncttrue
\mciteSetBstMidEndSepPunct{\mcitedefaultmidpunct}
{\mcitedefaultendpunct}{\mcitedefaultseppunct}\relax
\EndOfBibitem
\bibitem[Lewis \latin{et~al.}(2003)Lewis, hung Shih, Jones-Rhoades, Bartel, and
  Burge]{generic2}
Lewis,~B.~P.; hung Shih,~I.; Jones-Rhoades,~M.~W.; Bartel,~D.~P.; Burge,~C.~B.
  Prediction of Mammalian {MicroRNA} Targets. \emph{Cell} \textbf{2003},
  \emph{115}, 787--798\relax
\mciteBstWouldAddEndPuncttrue
\mciteSetBstMidEndSepPunct{\mcitedefaultmidpunct}
{\mcitedefaultendpunct}{\mcitedefaultseppunct}\relax
\EndOfBibitem
\bibitem[Wilson and Sharda(1994)Wilson, and Sharda]{generic3}
Wilson,~R.~L.; Sharda,~R. Bankruptcy prediction using neural networks.
  \emph{Decis. Support Syst.} \textbf{1994}, \emph{11}, 545--557\relax
\mciteBstWouldAddEndPuncttrue
\mciteSetBstMidEndSepPunct{\mcitedefaultmidpunct}
{\mcitedefaultendpunct}{\mcitedefaultseppunct}\relax
\EndOfBibitem
\bibitem[Geller(1997)]{generic4}
Geller,~R.~J. Earthquake prediction: a critical review. \emph{Geophys. J. Int.}
  \textbf{1997}, \emph{131}, 425--450\relax
\mciteBstWouldAddEndPuncttrue
\mciteSetBstMidEndSepPunct{\mcitedefaultmidpunct}
{\mcitedefaultendpunct}{\mcitedefaultseppunct}\relax
\EndOfBibitem
\bibitem[Becke(1993)]{becke3}
Becke,~A.~D. A new mixing of Hartree{\textendash}Fock and local
  density-functional theories. \emph{J. Chem. Phys.} \textbf{1993}, \emph{98},
  1372--1377\relax
\mciteBstWouldAddEndPuncttrue
\mciteSetBstMidEndSepPunct{\mcitedefaultmidpunct}
{\mcitedefaultendpunct}{\mcitedefaultseppunct}\relax
\EndOfBibitem
\bibitem[Lee \latin{et~al.}(1988)Lee, Yang, and Parr]{lyp}
Lee,~C.; Yang,~W.; Parr,~R.~G. Development of the Colle-Salvetti
  correlation-energy formula into a functional of the electron density.
  \emph{Phys. Rev. B} \textbf{1988}, \emph{37}, 785--789\relax
\mciteBstWouldAddEndPuncttrue
\mciteSetBstMidEndSepPunct{\mcitedefaultmidpunct}
{\mcitedefaultendpunct}{\mcitedefaultseppunct}\relax
\EndOfBibitem
\bibitem[Stephens \latin{et~al.}(1994)Stephens, Devlin, Chabalowski, and
  Frisch]{b3lyp}
Stephens,~P.~J.; Devlin,~F.~J.; Chabalowski,~C.~F.; Frisch,~M.~J. Ab Initio
  Calculation of Vibrational Absorption and Circular Dichroism Spectra Using
  Density Functional Force Fields. \emph{J. Phys. Chem.} \textbf{1994},
  \emph{98}, 11623--11627\relax
\mciteBstWouldAddEndPuncttrue
\mciteSetBstMidEndSepPunct{\mcitedefaultmidpunct}
{\mcitedefaultendpunct}{\mcitedefaultseppunct}\relax
\EndOfBibitem
\bibitem[Chai and Head-Gordon(2008)Chai, and Head-Gordon]{wB97XD}
Chai,~J.-D.; Head-Gordon,~M. Long-range corrected hybrid density functionals
  with damped atom{\textendash}atom dispersion corrections. \emph{Phys. Chem.
  Chem. Phys.} \textbf{2008}, \emph{10}, 6615\relax
\mciteBstWouldAddEndPuncttrue
\mciteSetBstMidEndSepPunct{\mcitedefaultmidpunct}
{\mcitedefaultendpunct}{\mcitedefaultseppunct}\relax
\EndOfBibitem
\bibitem[Elstner \latin{et~al.}(1998)Elstner, Porezag, Jungnickel, Elsner,
  Haugk, Frauenheim, Suhai, and Seifert]{dftb1}
Elstner,~M. \latin{et~al.}  Self-consistent-charge density-functional
  tight-binding method for simulations of complex materials properties.
  \emph{Phys. Rev. B} \textbf{1998}, \emph{58}, 7260--7268\relax
\mciteBstWouldAddEndPuncttrue
\mciteSetBstMidEndSepPunct{\mcitedefaultmidpunct}
{\mcitedefaultendpunct}{\mcitedefaultseppunct}\relax
\EndOfBibitem
\bibitem[Yang \latin{et~al.}(2007)Yang, Yu, York, Cui, and Elstner]{dftb2}
Yang,; Yu,~H.; York,~D.; Cui,~Q.; Elstner,~M. Extension of the
  Self-Consistent-Charge Density-Functional Tight-Binding Method: Third-Order
  Expansion of the Density Functional Theory Total Energy and Introduction of a
  Modified Effective Coulomb Interaction. \emph{J. Phys. Chem. A}
  \textbf{2007}, \emph{111}, 10861--10873\relax
\mciteBstWouldAddEndPuncttrue
\mciteSetBstMidEndSepPunct{\mcitedefaultmidpunct}
{\mcitedefaultendpunct}{\mcitedefaultseppunct}\relax
\EndOfBibitem
\bibitem[Gaus \latin{et~al.}(2011)Gaus, Cui, and Elstner]{dftb3}
Gaus,~M.; Cui,~Q.; Elstner,~M. DFTB3: Extension of the Self-Consistent-Charge
  Density-Functional Tight-Binding Method (SCC-DFTB). \emph{J. Chem. Theory
  Comput.} \textbf{2011}, \emph{7}, 931--948\relax
\mciteBstWouldAddEndPuncttrue
\mciteSetBstMidEndSepPunct{\mcitedefaultmidpunct}
{\mcitedefaultendpunct}{\mcitedefaultseppunct}\relax
\EndOfBibitem
\bibitem[Grimme \latin{et~al.}(2017)Grimme, Bannwarth, and Shushkov]{xtb1}
Grimme,~S.; Bannwarth,~C.; Shushkov,~P. A Robust and Accurate Tight-Binding
  Quantum Chemical Method for Structures, Vibrational Frequencies, and
  Noncovalent Interactions of Large Molecular Systems Parametrized for All
  spd-Block Elements (Z = 1–86). \emph{J. Chem. Theory Comput.}
  \textbf{2017}, \emph{13}, 1989--2009\relax
\mciteBstWouldAddEndPuncttrue
\mciteSetBstMidEndSepPunct{\mcitedefaultmidpunct}
{\mcitedefaultendpunct}{\mcitedefaultseppunct}\relax
\EndOfBibitem
\bibitem[Bannwarth \latin{et~al.}(2019)Bannwarth, Ehlert, and Grimme]{xtb2}
Bannwarth,~C.; Ehlert,~S.; Grimme,~S. GFN2-xTB—An Accurate and Broadly
  Parametrized Self-Consistent Tight-Binding Quantum Chemical Method with
  Multipole Electrostatics and Density-Dependent Dispersion Contributions.
  \emph{J. Chem. Theory Comput.} \textbf{2019}, \emph{15}, 1652--1671\relax
\mciteBstWouldAddEndPuncttrue
\mciteSetBstMidEndSepPunct{\mcitedefaultmidpunct}
{\mcitedefaultendpunct}{\mcitedefaultseppunct}\relax
\EndOfBibitem
\bibitem[Behler and Parrinello(2007)Behler, and Parrinello]{behler}
Behler,~J.; Parrinello,~M. Generalized Neural-Network Representation of
  High-Dimensional Potential-Energy Surfaces. \emph{Phys. Rev. Lett.}
  \textbf{2007}, \emph{98}, 146401\relax
\mciteBstWouldAddEndPuncttrue
\mciteSetBstMidEndSepPunct{\mcitedefaultmidpunct}
{\mcitedefaultendpunct}{\mcitedefaultseppunct}\relax
\EndOfBibitem
\bibitem[Schütt \latin{et~al.}(2018)Schütt, Sauceda, Kindermans, Tkatchenko,
  and Müller]{schnet}
Schütt,~K.~T.; Sauceda,~H.~E.; Kindermans,~P.-J.; Tkatchenko,~A.;
  Müller,~K.-R. SchNet – A deep learning architecture for molecules and
  materials. \emph{J. Chem. Phys.} \textbf{2018}, \emph{148}, 241722\relax
\mciteBstWouldAddEndPuncttrue
\mciteSetBstMidEndSepPunct{\mcitedefaultmidpunct}
{\mcitedefaultendpunct}{\mcitedefaultseppunct}\relax
\EndOfBibitem
\bibitem[Smith \latin{et~al.}(2018)Smith, Nebgen, Lubbers, Isayev, and
  Roitberg]{ani1x}
Smith,~J.~S.; Nebgen,~B.; Lubbers,~N.; Isayev,~O.; Roitberg,~A.~E. Less is
  more: Sampling chemical space with active learning. \emph{J. Chem. Phys.}
  \textbf{2018}, \emph{148}, 241733\relax
\mciteBstWouldAddEndPuncttrue
\mciteSetBstMidEndSepPunct{\mcitedefaultmidpunct}
{\mcitedefaultendpunct}{\mcitedefaultseppunct}\relax
\EndOfBibitem
\bibitem[Shao \latin{et~al.}(2020)Shao, Hellstr\"{o}m, Mitev, Knijff, and
  Zhang]{pinn}
Shao,~Y.; Hellstr\"{o}m,~M.; Mitev,~P.~D.; Knijff,~L.; Zhang,~C. {PiNN}: A
  Python Library for Building Atomic Neural Networks of Molecules and
  Materials. \emph{J. Chem. Inf. Model.} \textbf{2020}, \emph{60},
  1184--1193\relax
\mciteBstWouldAddEndPuncttrue
\mciteSetBstMidEndSepPunct{\mcitedefaultmidpunct}
{\mcitedefaultendpunct}{\mcitedefaultseppunct}\relax
\EndOfBibitem
\bibitem[Cornell \latin{et~al.}(1995)Cornell, Cieplak, Bayly, Gould, Merz,
  Ferguson, Spellmeyer, Fox, Caldwell, and Kollman]{amber}
Cornell,~W.~D. \latin{et~al.}  A Second Generation Force Field for the
  Simulation of Proteins, Nucleic Acids, and Organic Molecules. \emph{J. Am.
  Chem. Soc.} \textbf{1995}, \emph{117}, 5179--5197\relax
\mciteBstWouldAddEndPuncttrue
\mciteSetBstMidEndSepPunct{\mcitedefaultmidpunct}
{\mcitedefaultendpunct}{\mcitedefaultseppunct}\relax
\EndOfBibitem
\bibitem[MacKerell \latin{et~al.}(1998)MacKerell, Bashford, Bellott, Dunbrack,
  Evanseck, Field, Fischer, Gao, Guo, Ha, Joseph-McCarthy, Kuchnir, Kuczera,
  Lau, Mattos, Michnick, Ngo, Nguyen, Prodhom, Reiher, Roux, Schlenkrich,
  Smith, Stote, Straub, Watanabe, Wiórkiewicz-Kuczera, Yin, and
  Karplus]{charmm}
MacKerell,~A.~D. \latin{et~al.}  All-Atom Empirical Potential for Molecular
  Modeling and Dynamics Studies of Proteins. \emph{J. Phys. Chem. B}
  \textbf{1998}, \emph{102}, 3586--3616\relax
\mciteBstWouldAddEndPuncttrue
\mciteSetBstMidEndSepPunct{\mcitedefaultmidpunct}
{\mcitedefaultendpunct}{\mcitedefaultseppunct}\relax
\EndOfBibitem
\bibitem[van Duin \latin{et~al.}(2001)van Duin, Dasgupta, Lorant, and
  Goddard]{reax1}
van Duin,~A. C.~T.; Dasgupta,~S.; Lorant,~F.; Goddard,~W.~A. ReaxFF: A Reactive
  Force Field for Hydrocarbons. \emph{J. Phys. Chem. A} \textbf{2001},
  \emph{105}, 9396--9409\relax
\mciteBstWouldAddEndPuncttrue
\mciteSetBstMidEndSepPunct{\mcitedefaultmidpunct}
{\mcitedefaultendpunct}{\mcitedefaultseppunct}\relax
\EndOfBibitem
\bibitem[Chenoweth \latin{et~al.}(2008)Chenoweth, van Duin, and Goddard]{reax2}
Chenoweth,~K.; van Duin,~A. C.~T.; Goddard,~W.~A. ReaxFF Reactive Force Field
  for Molecular Dynamics Simulations of Hydrocarbon Oxidation. \emph{J. Phys.
  Chem. A} \textbf{2008}, \emph{112}, 1040--1053\relax
\mciteBstWouldAddEndPuncttrue
\mciteSetBstMidEndSepPunct{\mcitedefaultmidpunct}
{\mcitedefaultendpunct}{\mcitedefaultseppunct}\relax
\EndOfBibitem
\bibitem[Smith \latin{et~al.}(2019)Smith, Nebgen, Zubatyuk, Lubbers, Devereux,
  Barros, Tretiak, Isayev, and Roitberg]{ani1ccx}
Smith,~J.~S. \latin{et~al.}  {Approaching coupled cluster accuracy with a
  general-purpose neural network potential through transfer learning}. Preprint
  at
  \url{https://chemrxiv.org/articles/Outsmarting_Quantum_Chemistry_Through_Transfer_Learning/6744440},
  2019\relax
\mciteBstWouldAddEndPuncttrue
\mciteSetBstMidEndSepPunct{\mcitedefaultmidpunct}
{\mcitedefaultendpunct}{\mcitedefaultseppunct}\relax
\EndOfBibitem
\bibitem[Bannwarth \latin{et~al.}(2019)Bannwarth, Ehlert, and Grimme]{gfn2xtb}
Bannwarth,~C.; Ehlert,~S.; Grimme,~S. {GFN}2-{xTB}{\textemdash}An Accurate and
  Broadly Parametrized Self-Consistent Tight-Binding Quantum Chemical Method
  with Multipole Electrostatics and Density-Dependent Dispersion Contributions.
  \emph{J. Chem. Theory Comput.} \textbf{2019}, \emph{15}, 1652--1671\relax
\mciteBstWouldAddEndPuncttrue
\mciteSetBstMidEndSepPunct{\mcitedefaultmidpunct}
{\mcitedefaultendpunct}{\mcitedefaultseppunct}\relax
\EndOfBibitem
\bibitem[Shchygol \latin{et~al.}(2019)Shchygol, Yakovlev, Trnka, van Duin, and
  Verstraelen]{anna}
Shchygol,~G.; Yakovlev,~A.; Trnka,~T.; van Duin,~A. C.~T.; Verstraelen,~T.
  ReaxFF Parameter Optimization with Monte-Carlo and Evolutionary Algorithms:
  Guidelines and Insights. \emph{J. Chem. Theory Comput.} \textbf{2019},
  \emph{15}, 6799--6812\relax
\mciteBstWouldAddEndPuncttrue
\mciteSetBstMidEndSepPunct{\mcitedefaultmidpunct}
{\mcitedefaultendpunct}{\mcitedefaultseppunct}\relax
\EndOfBibitem
\bibitem[Dieterich and Hartke(2010)Dieterich, and Hartke]{ogolem}
Dieterich,~J.~M.; Hartke,~B. OGOLEM: Global cluster structure optimisation for
  arbitrary mixtures of flexible molecules. A multiscaling, object-oriented
  approach. \emph{Mol. Phys.} \textbf{2010}, \emph{108}, 279--291\relax
\mciteBstWouldAddEndPuncttrue
\mciteSetBstMidEndSepPunct{\mcitedefaultmidpunct}
{\mcitedefaultendpunct}{\mcitedefaultseppunct}\relax
\EndOfBibitem
\bibitem[Wang \latin{et~al.}(2012)Wang, Chen, and Voorhis]{forcebalance}
Wang,~L.-P.; Chen,~J.; Voorhis,~T.~V. Systematic Parametrization of Polarizable
  Force Fields from Quantum Chemistry Data. \emph{J. Chem. Theory Comput.}
  \textbf{2012}, \emph{9}, 452--460\relax
\mciteBstWouldAddEndPuncttrue
\mciteSetBstMidEndSepPunct{\mcitedefaultmidpunct}
{\mcitedefaultendpunct}{\mcitedefaultseppunct}\relax
\EndOfBibitem
\bibitem[Brommer \latin{et~al.}(2015)Brommer, Kiselev, Schopf, Beck, Roth, and
  Trebin]{potfit}
Brommer,~P. \latin{et~al.}  Classical interaction potentials for diverse
  materials fromab initiodata: a review ofpotfit. \emph{Model. Simul. Mater.
  Sc.} \textbf{2015}, \emph{23}, 074002\relax
\mciteBstWouldAddEndPuncttrue
\mciteSetBstMidEndSepPunct{\mcitedefaultmidpunct}
{\mcitedefaultendpunct}{\mcitedefaultseppunct}\relax
\EndOfBibitem
\bibitem[Jaramillo-Botero \latin{et~al.}(2014)Jaramillo-Botero, Naserifar, and
  Goddard]{garfield}
Jaramillo-Botero,~A.; Naserifar,~S.; Goddard,~W.~A. General Multiobjective
  Force Field Optimization Framework, with Application to Reactive Force Fields
  for Silicon Carbide. \emph{J. Chem. Theory Comput.} \textbf{2014}, \emph{10},
  1426--1439\relax
\mciteBstWouldAddEndPuncttrue
\mciteSetBstMidEndSepPunct{\mcitedefaultmidpunct}
{\mcitedefaultendpunct}{\mcitedefaultseppunct}\relax
\EndOfBibitem
\bibitem[Senftle \latin{et~al.}(2016)Senftle, Hong, Islam, Kylasa, Zheng, Shin,
  Junkermeier, Engel-Herbert, Janik, Aktulga, Verstraelen, Grama, and van
  Duin]{reax_npj}
Senftle,~T.~P. \latin{et~al.}  The ReaxFF reactive force-field: development,
  applications and future directions. \emph{npj Comput. Mat.} \textbf{2016},
  \emph{2}, 15011\relax
\mciteBstWouldAddEndPuncttrue
\mciteSetBstMidEndSepPunct{\mcitedefaultmidpunct}
{\mcitedefaultendpunct}{\mcitedefaultseppunct}\relax
\EndOfBibitem
\bibitem[Guvench \latin{et~al.}(2009)Guvench, Hatcher, Venable, Pastor, and
  MacKerell]{fitting1}
Guvench,~O.; Hatcher,~E.; Venable,~R.~M.; Pastor,~R.~W.; MacKerell,~A.~D.
  CHARMM Additive All-Atom Force Field for Glycosidic Linkages between
  Hexopyranoses. \emph{J. Chem. Theory Comput.} \textbf{2009}, \emph{5},
  2353--2370\relax
\mciteBstWouldAddEndPuncttrue
\mciteSetBstMidEndSepPunct{\mcitedefaultmidpunct}
{\mcitedefaultendpunct}{\mcitedefaultseppunct}\relax
\EndOfBibitem
\bibitem[Gaus \latin{et~al.}(2009)Gaus, Chou, Witek, and Elstner]{fitting2}
Gaus,~M.; Chou,~C.-P.; Witek,~H.; Elstner,~M. Automatized Parametrization of
  SCC-DFTB Repulsive Potentials: Application to Hydrocarbons. \emph{J. Phys.
  Chem. A} \textbf{2009}, \emph{113}, 11866--11881\relax
\mciteBstWouldAddEndPuncttrue
\mciteSetBstMidEndSepPunct{\mcitedefaultmidpunct}
{\mcitedefaultendpunct}{\mcitedefaultseppunct}\relax
\EndOfBibitem
\bibitem[van Beest \latin{et~al.}(1990)van Beest, Kramer, and van
  Santen]{fitting3}
van Beest,~B. W.~H.; Kramer,~G.~J.; van Santen,~R.~A. Force fields for silicas
  and aluminophosphates based on ab initio calculations. \emph{Phys. Rev.
  Lett.} \textbf{1990}, \emph{64}, 1955--1958\relax
\mciteBstWouldAddEndPuncttrue
\mciteSetBstMidEndSepPunct{\mcitedefaultmidpunct}
{\mcitedefaultendpunct}{\mcitedefaultseppunct}\relax
\EndOfBibitem
\bibitem[Ashraf and van Duin(2017)Ashraf, and van Duin]{fitting4}
Ashraf,~C.; van Duin,~A.~C. Extension of the ReaxFF Combustion Force Field
  toward Syngas Combustion and Initial Oxidation Kinetics. \emph{J. Phys. Chem.
  A} \textbf{2017}, \emph{121}, 1051--1068\relax
\mciteBstWouldAddEndPuncttrue
\mciteSetBstMidEndSepPunct{\mcitedefaultmidpunct}
{\mcitedefaultendpunct}{\mcitedefaultseppunct}\relax
\EndOfBibitem
\bibitem[van Duin(2002)]{reax_manual}
van Duin,~A. C.~T. {ReaxFF User Manual}.
  \url{https://www.scm.com/doc/ReaxFF/index.html}, 2002\relax
\mciteBstWouldAddEndPuncttrue
\mciteSetBstMidEndSepPunct{\mcitedefaultmidpunct}
{\mcitedefaultendpunct}{\mcitedefaultseppunct}\relax
\EndOfBibitem
\bibitem[Wang()]{forcebalance_manual}
Wang,~L.-P. ForceBalance: Main Page.
  \url{http://leeping.github.io/forcebalance/doc/html/index.html},
  \url{http://leeping.github.io/forcebalance/doc/html/index.html}\relax
\mciteBstWouldAddEndPuncttrue
\mciteSetBstMidEndSepPunct{\mcitedefaultmidpunct}
{\mcitedefaultendpunct}{\mcitedefaultseppunct}\relax
\EndOfBibitem
\bibitem[Dieterich and Hartke()Dieterich, and Hartke]{ogolem_manual}
Dieterich,~J.~M.; Hartke,~B. OGOLEM.ORG Homepage.
  \url{{https://www.ogolem.org/manual/}},
  \url{https://www.ogolem.org/manual/}\relax
\mciteBstWouldAddEndPuncttrue
\mciteSetBstMidEndSepPunct{\mcitedefaultmidpunct}
{\mcitedefaultendpunct}{\mcitedefaultseppunct}\relax
\EndOfBibitem
\bibitem[Martinez \latin{et~al.}(2016)Martinez, Chernatynskiy, Yilmaz, Liang,
  Sinnott, and Phillpot]{posmat}
Martinez,~J.~A. \latin{et~al.}  Potential Optimization Software for Materials
  ({POSMat}). \emph{Comput. Phys. Commun.} \textbf{2016}, \emph{203},
  201--211\relax
\mciteBstWouldAddEndPuncttrue
\mciteSetBstMidEndSepPunct{\mcitedefaultmidpunct}
{\mcitedefaultendpunct}{\mcitedefaultseppunct}\relax
\EndOfBibitem
\bibitem[Komissarov and Rüger()Komissarov, and Rüger]{doc}
Komissarov,~L.; Rüger,~R. ParAMS Documentation.
  \url{https://www.scm.com/doc.trunk/params/index.html},
  \url{https://www.scm.com/doc.trunk/params/index.html}\relax
\mciteBstWouldAddEndPuncttrue
\mciteSetBstMidEndSepPunct{\mcitedefaultmidpunct}
{\mcitedefaultendpunct}{\mcitedefaultseppunct}\relax
\EndOfBibitem
\bibitem[Komissarov \latin{et~al.}(2020)Komissarov, Rüger, Hellström, and
  Verstraelen]{si}
Komissarov,~L.; Rüger,~R.; Hellström,~M.; Verstraelen,~T. ParAMS Supporting
  Information. \url{www.doi.org/10.5281/zenodo.4629706}, 2020\relax
\mciteBstWouldAddEndPuncttrue
\mciteSetBstMidEndSepPunct{\mcitedefaultmidpunct}
{\mcitedefaultendpunct}{\mcitedefaultseppunct}\relax
\EndOfBibitem
\bibitem[döt Net \latin{et~al.}(2019)döt Net, Evans, and Ben-Kiki]{yaml}
döt Net,~I.; Evans,~C.; Ben-Kiki,~O. The Official YAML Web Site.
  \url{https://yaml.org/}, 2019\relax
\mciteBstWouldAddEndPuncttrue
\mciteSetBstMidEndSepPunct{\mcitedefaultmidpunct}
{\mcitedefaultendpunct}{\mcitedefaultseppunct}\relax
\EndOfBibitem
\bibitem[Rüger \latin{et~al.}(2019)Rüger, Franchini, Trnka, Yakovlev, van
  Lenthe, Philipsen, van Vuren, Klumpers, and Soini]{ams}
Rüger, \latin{et~al.}  Amsterdam Modeling Suite. \url{https://scm.com},
  2019\relax
\mciteBstWouldAddEndPuncttrue
\mciteSetBstMidEndSepPunct{\mcitedefaultmidpunct}
{\mcitedefaultendpunct}{\mcitedefaultseppunct}\relax
\EndOfBibitem
\bibitem[van Duin \latin{et~al.}(2019)van Duin, Goddard, Islam, van Schoot,
  Trnka, and Yakovlev]{scm}
van Duin, \latin{et~al.}  ReaxFF 2019.4. \url{https://www.scm.com/}, 2019\relax
\mciteBstWouldAddEndPuncttrue
\mciteSetBstMidEndSepPunct{\mcitedefaultmidpunct}
{\mcitedefaultendpunct}{\mcitedefaultseppunct}\relax
\EndOfBibitem
\bibitem[Hansen and Ostermeier(2001)Hansen, and Ostermeier]{cma1}
Hansen,~N.; Ostermeier,~A. Completely Derandomized Self-Adaptation in Evolution
  Strategies. \emph{Evol. Comput.} \textbf{2001}, \emph{9}, 159--195\relax
\mciteBstWouldAddEndPuncttrue
\mciteSetBstMidEndSepPunct{\mcitedefaultmidpunct}
{\mcitedefaultendpunct}{\mcitedefaultseppunct}\relax
\EndOfBibitem
\bibitem[Hansen and Kern(2004)Hansen, and Kern]{cma2}
Hansen,~N.; Kern,~S. Evaluating the {CMA} Evolution Strategy on Multimodal Test
  Functions. Parallel Problem Solving from Nature {PPSN VIII}. 2004; pp
  282--291\relax
\mciteBstWouldAddEndPuncttrue
\mciteSetBstMidEndSepPunct{\mcitedefaultmidpunct}
{\mcitedefaultendpunct}{\mcitedefaultseppunct}\relax
\EndOfBibitem
\bibitem[Hansen \latin{et~al.}(2019)Hansen, Akimoto, and Baudis]{cma_py}
Hansen,~N.; Akimoto,~Y.; Baudis,~P. {CMA-ES/pycma} on {G}ithub.
  \url{https://doi.org/10.5281/zenodo.2559634}, 2019;
  \url{https://doi.org/10.5281/zenodo.2559634}\relax
\mciteBstWouldAddEndPuncttrue
\mciteSetBstMidEndSepPunct{\mcitedefaultmidpunct}
{\mcitedefaultendpunct}{\mcitedefaultseppunct}\relax
\EndOfBibitem
\bibitem[Cosseddu and Infante(2016)Cosseddu, and Infante]{armc}
Cosseddu,~S.; Infante,~I. Force Field Parametrization of Colloidal {CdSe}
  Nanocrystals Using an Adaptive Rate Monte Carlo Optimization Algorithm.
  \emph{J. Chem. Theory Comput.} \textbf{2016}, \emph{13}, 297--308\relax
\mciteBstWouldAddEndPuncttrue
\mciteSetBstMidEndSepPunct{\mcitedefaultmidpunct}
{\mcitedefaultendpunct}{\mcitedefaultseppunct}\relax
\EndOfBibitem
\bibitem[Rapin and Teytaud(2018)Rapin, and Teytaud]{nevergrad}
Rapin,~J.; Teytaud,~O. {Nevergrad - A gradient-free optimization platform}.
  \url{https://GitHub.com/FacebookResearch/Nevergrad}, 2018\relax
\mciteBstWouldAddEndPuncttrue
\mciteSetBstMidEndSepPunct{\mcitedefaultmidpunct}
{\mcitedefaultendpunct}{\mcitedefaultseppunct}\relax
\EndOfBibitem
\bibitem[{Virtanen} \latin{et~al.}(2020){Virtanen}, {Gommers}, {Oliphant},
  {Haberland}, {Reddy}, {Cournapeau}, {Burovski}, {Peterson}, {Weckesser},
  {Bright}, {van der Walt}, {Brett}, {Wilson}, {Jarrod Millman}, {Mayorov},
  {Nelson}, {Jones}, {Kern}, {Larson}, {Carey}, {Polat}, {Feng}, {Moore}, {Vand
  erPlas}, {Laxalde}, {Perktold}, {Cimrman}, {Henriksen}, {Quintero}, {Harris},
  {Archibald}, {Ribeiro}, {Pedregosa}, {van Mulbregt}, and
  {Contributors}]{scipy}
{Virtanen},~P. \latin{et~al.}  {SciPy 1.0: Fundamental Algorithms for
  Scientific Computing in Python}. \emph{Nat. Methods} \textbf{2020},
  \emph{17}, 261--272\relax
\mciteBstWouldAddEndPuncttrue
\mciteSetBstMidEndSepPunct{\mcitedefaultmidpunct}
{\mcitedefaultendpunct}{\mcitedefaultseppunct}\relax
\EndOfBibitem
\bibitem[Kamat \latin{et~al.}(2010)Kamat, van Duin, and Yakovlev]{reax_ADF}
Kamat,~A.~M.; van Duin,~A. C.~T.; Yakovlev,~A. Molecular Dynamics Simulations
  of Laser-Induced Incandescence of Soot Using an Extended ReaxFF Reactive
  Force Field. \emph{J. Phys. Chem. A} \textbf{2010}, \emph{114},
  12561--12572\relax
\mciteBstWouldAddEndPuncttrue
\mciteSetBstMidEndSepPunct{\mcitedefaultmidpunct}
{\mcitedefaultendpunct}{\mcitedefaultseppunct}\relax
\EndOfBibitem
\bibitem[Hunter(2007)]{matplotlib}
Hunter,~J.~D. Matplotlib: A 2D graphics environment. \emph{Comput. Sci. Eng.}
  \textbf{2007}, \emph{9}, 90--95\relax
\mciteBstWouldAddEndPuncttrue
\mciteSetBstMidEndSepPunct{\mcitedefaultmidpunct}
{\mcitedefaultendpunct}{\mcitedefaultseppunct}\relax
\EndOfBibitem
\bibitem[Hellström \latin{et~al.}(2013)Hellström, Jorner, Bryngelsson, Huber,
  Kullgren, Frauenheim, and Broqvist]{znopt}
Hellström,~M. \latin{et~al.}  An SCC-DFTB Repulsive Potential for Various ZnO
  Polymorphs and the ZnO–Water System. \emph{J. Phys. Chem. C} \textbf{2013},
  \emph{117}, 17004--17015\relax
\mciteBstWouldAddEndPuncttrue
\mciteSetBstMidEndSepPunct{\mcitedefaultmidpunct}
{\mcitedefaultendpunct}{\mcitedefaultseppunct}\relax
\EndOfBibitem
\bibitem[Moreira \latin{et~al.}(2009)Moreira, Dolgonos, Aradi, da~Rosa, and
  Frauenheim]{znorg}
Moreira,~N.~H.; Dolgonos,~G.; Aradi,~B.; da~Rosa,~A.~L.; Frauenheim,~T. Toward
  an Accurate Density-Functional Tight-Binding Description of Zinc-Containing
  Compounds. \emph{J. Chem. Theory Comput.} \textbf{2009}, \emph{5},
  605--614\relax
\mciteBstWouldAddEndPuncttrue
\mciteSetBstMidEndSepPunct{\mcitedefaultmidpunct}
{\mcitedefaultendpunct}{\mcitedefaultseppunct}\relax
\EndOfBibitem
\bibitem[te~Velde and Baerends(1991)te~Velde, and Baerends]{band}
te~Velde,~G.; Baerends,~E.~J. Precise density-functional method for periodic
  structures. \emph{Phys. Rev. B} \textbf{1991}, \emph{44}, 7888--7903\relax
\mciteBstWouldAddEndPuncttrue
\mciteSetBstMidEndSepPunct{\mcitedefaultmidpunct}
{\mcitedefaultendpunct}{\mcitedefaultseppunct}\relax
\EndOfBibitem
\bibitem[Nelder and Mead(1965)Nelder, and Mead]{neldermead}
Nelder,~J.~A.; Mead,~R. A Simplex Method for Function Minimization.
  \emph{Comput. J.} \textbf{1965}, \emph{7}, 308--313\relax
\mciteBstWouldAddEndPuncttrue
\mciteSetBstMidEndSepPunct{\mcitedefaultmidpunct}
{\mcitedefaultendpunct}{\mcitedefaultseppunct}\relax
\EndOfBibitem
\bibitem[Strachan \latin{et~al.}(2003)Strachan, van Duin, Chakraborty,
  Dasgupta, and Goddard]{reax_RDX}
Strachan,~A.; van Duin,~A. C.~T.; Chakraborty,~D.; Dasgupta,~S.; Goddard,~W.~A.
  Shock Waves in High-Energy Materials: The Initial Chemical Events in
  Nitramine RDX. \emph{Phys. Rev. Lett.} \textbf{2003}, \emph{91}, 098301\relax
\mciteBstWouldAddEndPuncttrue
\mciteSetBstMidEndSepPunct{\mcitedefaultmidpunct}
{\mcitedefaultendpunct}{\mcitedefaultseppunct}\relax
\EndOfBibitem
\bibitem[Fogarty \latin{et~al.}(2010)Fogarty, Aktulga, Grama, van Duin, and
  Pandit]{reax_Silica}
Fogarty,~J.~C.; Aktulga,~H.~M.; Grama,~A.~Y.; van Duin,~A. C.~T.; Pandit,~S.~A.
  A reactive molecular dynamics simulation of the silica-water interface.
  \emph{J. Chem. Phys.} \textbf{2010}, \emph{132}, 174704\relax
\mciteBstWouldAddEndPuncttrue
\mciteSetBstMidEndSepPunct{\mcitedefaultmidpunct}
{\mcitedefaultendpunct}{\mcitedefaultseppunct}\relax
\EndOfBibitem
\bibitem[Müller and Hartke(2016)Müller, and Hartke]{muellerhartke}
Müller,~J.; Hartke,~B. reaxFF Reactive Force Field for Disulfide
  Mechanochemistry, Fitted to Multireference ab Initio Data. \emph{J. Chem.
  Theory Comput.} \textbf{2016}, \emph{12}, 3913--3925\relax
\mciteBstWouldAddEndPuncttrue
\mciteSetBstMidEndSepPunct{\mcitedefaultmidpunct}
{\mcitedefaultendpunct}{\mcitedefaultseppunct}\relax
\EndOfBibitem
\bibitem[Furman and Wales(2019)Furman, and Wales]{tapered_bond_orders}
Furman,~D.; Wales,~D.~J. Transforming the Accuracy and Numerical Stability of
  ReaxFF Reactive Force Fields. \emph{J. Phys. Chem. Lett.} \textbf{2019},
  \emph{10}, 7215--7223\relax
\mciteBstWouldAddEndPuncttrue
\mciteSetBstMidEndSepPunct{\mcitedefaultmidpunct}
{\mcitedefaultendpunct}{\mcitedefaultseppunct}\relax
\EndOfBibitem
\end{mcitethebibliography}
}

\begin{tocentry}
\includegraphics[width=3.5in]{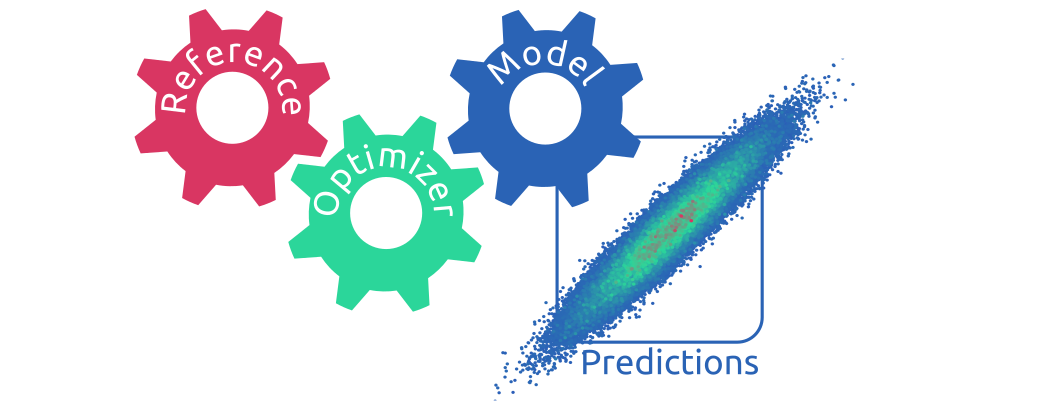}
\end{tocentry}

\end{document}


\maketitle

\section{The Parameterization Problem}

This section provides a mathematical framework for the parameterization problem. 
We assume that the training data can be defined as a set of physico-chemical properties for a number of isolated or periodic systems.
Examples for relevant properties are energy differences, nuclear gradients or system geometries.
In the context of ParAMS, we define an arbitrary property $P$ 
that can be expressed as the output of a computational job.
When working with multiple jobs as part of a training set, a job function
can be defined as

\begin{equation}
    \label{eq:jobcollection} 
    J(R_j, S_j, M) =
        \left( R'_j, P_j^n\right)
        \quad \forall
            \; j\in \{1\ldots N_\text{job}\},
            \; n\in \{1\ldots N_\text{prop}(j)\},
\end{equation}

calculating the output geometry $R'_j$ and all properties $P_j^n$ of a job $j$.
The input for every job in $J$ consists of the input geometry $R_j$, 
the job settings $S_j$ (\eg geometry optimization and frequencies)
and the computational model $M$.
Note that a parametric model is additionally a function of the parameter vector $\bm{x}$,
in which case the outputs of the above equation can be denoted with the hat operator
(\ie $ \hat{R}'_j, \hat{P}_j^n$),
as to distinguish between reference properties and properties predicted by the parametric model.
Training set entries can be constructed, for example, from a linear combination of multiple properties
\begin{align}
    \label{eq:dsentry}
    y_i &=
        \sum_{k=1}^{N_\text{lc}(i)} c_{i,k} P_{j(i,k)}^{n(i,k)}
        \quad \forall\; i\in \{1\ldots N_\text{data}\},
\end{align}
where $c_{i,k}$ is the coefficient for term $k$ of training set entry $i$ and $N_\text{lc}(i)$ is the total number of terms
per entry.
Non-linear combinations of properties to construct $y_i$ are also possible.
Such a formulation offers a high degree of flexibility for the construction of a training set.
One example is the combination of multiple system energies into one reaction energy.
It should be noted that a training set entry, as defined in Eq.\ \ref{eq:dsentry},
does not have to originate from the results of computational jobs.
The reference value can instead be provided directly, making it easy to work with experimental or external data.

While training set entries $\bm{y}$ have to be defined only once,
their predicted counterpart $\bm{\hat{y}}$ has to be re-calculated every time the model parameters
change.
For this purpose, we introduce a Data Set function
\begin{equation}
    \label{eq:dataset}
    DS(\bm{x} | \bm{y}) = \bm{y} - \hat{\bm{y}},
\end{equation}
which extracts all properties needed for the calculation of $\bm{\hat{y}}$
based on a parameter set $\bm{x}$ and returns the respective vector of residuals.
A metric in the form of a loss function $L(\bm{y} - \hat{\bm{y}}, \bm{w})$
is then applied to the residuals for a qualitative measure of how close
reference and predicted values  are.
The additional weights vector vector $\bm{w}$ can be used to balance
possibly different orders of magnitude in the data set or
make certain entries more relevant for the fitting process than others.

Finally, the optimization algorithm can be defined as a function that minimizes $L$ with respect to the parameters
\begin{equation}
    \label{eq:optimizer}
    O(\bm{x}_0, L) = 
    \underset{\bm{x}} {\mathop{\mathrm{arg\,min}}}\,
        L = \bm{x}^*,
\end{equation}
finding an optimal solution $\bm{x}^*$ from an initial point $\bm{x}_0$.

\section{Additional Display Items}
\begin{table}
\centering
\caption{
Composition of the reference data published by M\"uller and Hartke\cite{muellerhartke},
split by the computational tasks Single Point (SP) and Geometry Optimization (GO).
For each of the two sets, the upper part describes the chemical systems,
while the lower breaks down the individual entries in the training and validation sets.
Note that some entries might be a function of multiple chemical systems,
meaning that the sum of SP+GO is not necessarily equal to the total
number of entries for that row (\cf Sec.\ 3.1 in the main text).
}
\begin{tabular}{lrrr}

    \textbf{Training Set}&
    \textbf{SP}&
    \textbf{GO}&
    \textbf{Total}
    \\
    \hline
    
    Number of systems&
    222&
    9&
    231
    \\
    
    Mean system size (atoms)&
    6.6&
    11.4&
    6.8
    \\
    
    Std. dev. (atoms)&
    2.9&
    7.7&
    3.3
    \\

    \\
    Total number of entries&
    4620&
    317&
    4875
    \\
    
    Energies&
    219&
    62&
    219
    \\
    
    Forces&
    4401&
    0&
    4401
    \\
    
    Atomic distances&
    0&
    94&
    94
    \\
    
    Angles&
    0&
    85&
    85
    \\
    
    Dihedrals&
    0&
    76&
    76
    \\
   

    \\
    \textbf{Validation Set}&
    &
    &
    \\
    \hline
    
    Number of systems&
    200&
    24&
    224
    \\
    
    Mean system size (atoms)&
    24.0&
    12.7&
    22.8
    \\
    
    Std. dev. (atoms)&
    0.0&
    5.9&
    4.0
    \\

    \\    
    Total number of entries&
    199&
    771&
    970
    \\
    
    Energies&
    199&
    0&
    199
    \\
    
    Forces&
    &
    &
    0
    \\
    
    Atomic distances&
    0&
    281&
    281
    \\
    
    Angles&
    0&
    257&
    257
    \\
    
    Dihedrals&
    0&
    233&
    233
\label{tab:trainingset}
\end{tabular}
\end{table}

\begin{table}
\centering
\caption{Summary of relevant ParAMS settings used for the re-parameterization of Mue2016.}
\begin{tabular}{lr}
    \textbf{Setting}&
    \textbf{Value}
    \\
    \hline

    Number of optimizations&
    9
    \\
    
    Number of parameters to optimize&
    35
    \\
    
    Lower / upper parameter bounds&
    $\bm{x}_0 \pm 0.2|\bm{x}_0|$
    \\
    
    Optimization timeout&
    24 hours
    \\
    
    CMA-ES population size&
    36
    \\
    
    CMA-ES sigma&
    0.3
    \\
    
    Loss function&
    sum of squared errors
    \\
    
    Early stopping patience&
    6000 evaluations
    \\
    
    Constraints&
    $r_0^\sigma \geq r_0^\pi$
    and
    $r_0^\pi \geq r_0^{\pi\pi}$
\label{tab:params_setup}
\end{tabular}
\end{table}

\begin{figure}
  \centering
  \includegraphics[width=.45\textwidth]{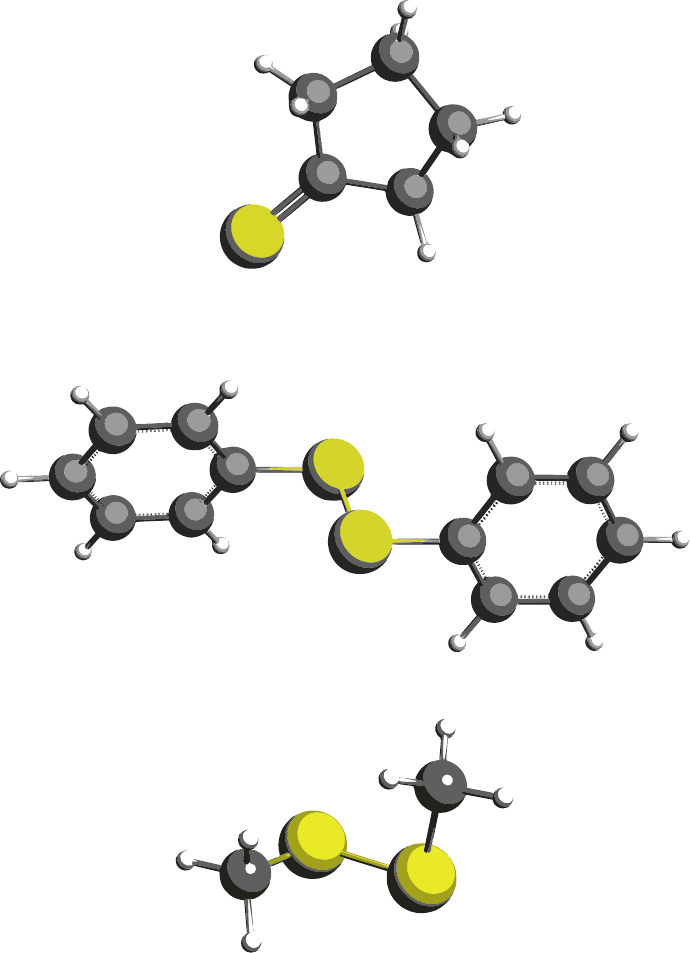}
  \caption{
    From top to bottom: Example structures of
    cyclopentathione, diphenyl disulfide and dimethyl disulfide, 
    containing S (yellow), C (black), and H (white),
    included in the data provided by MH.
    The fitted properties include bond distances, angles, relative
    energies and atomic forces.
    }
  \label{fig:compounds}
\end{figure}

\newpage
\bibliography{references}